\documentclass[aps,pra,twocolumn,showpacs,superscriptaddress,floatfix,nofootinbib]{revtex4-1}

\usepackage{graphicx}
\usepackage{amsmath}
\usepackage{epsfig}
\usepackage{helvet}
\usepackage{amssymb}

\newcommand{\be}{\begin{equation}}
\newcommand{\ee}{\end{equation}}
\newcommand{\bea}{\begin{eqnarray}}
\newcommand{\eea}{\end{eqnarray}}

\newcommand{\bra}[1]{\mbox{$\langle #1 |$}}
\newcommand{\ket}[1]{\mbox{$| #1 \rangle$}}

\def\PsiId{\Psi_{\mbox{\tiny 1}}}
\def\PsiSe{\Psi_{\mbox{\tiny s}}}

\def\sss{\mbox{\tiny s}}

\newcommand{\Tor}{\mbox{\tiny tor}}
\newcommand{\Cyl}{\mbox{\tiny cyl}}

\newcommand{\HH}{H_{\mbox{\tiny Hal.}}}

\begin{document}

\title{
Characterizing topological order by studying the ground states of an infinite cylinder
}
\author{L. Cincio}
\author{G. Vidal}
\affiliation{Perimeter Institute for Theoretical Physics, Waterloo, Ontario, N2L 2Y5, Canada}
\date{\today}

\begin{abstract}
Given a microscopic lattice Hamiltonian for a topologically ordered phase, we describe a tensor network approach to characterize its emergent anyon model and, in a chiral phase, also its gapless edge theory.
First, a tensor network representation of a complete, orthonormal set of ground states on a cylinder of infinite length and finite width is obtained through numerical optimization. Each of these ground states is argued to have a different anyonic flux threading through the cylinder. In a chiral phase, the entanglement spectrum of each ground state is seen to reveal a different sector of the corresponding gapless edge theory.
A quasi-orthogonal basis on the torus is then produced by chopping off and reconnecting the tensor network representation on the cylinder. Elaborating on the recent proposal of [Y. Zhang et al. Phys. Rev. B 85, 235151 (2012)], a rotation on the torus yields an alternative basis of ground states and, through the computation of overlaps between bases, the modular matrices $S$ and $U$ (containing the mutual and self statistics of the different anyon species) are extracted. As an application, we study the hard-core boson Haldane model by using the two-dimensional density matrix renormalization group. A thorough characterization of the universal properties of this lattice model, both in the bulk and at the edge, unambiguously shows that its ground space realizes the $\nu=1/2$ bosonic Laughlin state.
\end{abstract}
%\pacs{03.67.-a, 03.65.Ud, 02.70.-c, 05.30.Fk}

\maketitle

For many decades, characterizing the emergent order of an interacting quantum many-body system from its microscopic description has been regarded as an extremely challenging task. Consider, for instance, a lattice Hamiltonian $H$ suspected of realizing some form of topological order \cite{Wen90}, say a given Laughlin state \cite{Laughlin83} or a quantum spin liquid \cite{Anderson73, Balents10} ---states of considerable interest both in the study of exotic many-body phenomena such as the fractional quantum Hall (FQH) effect \cite{Tsui82} or high-temperature superconductivity \cite{Bednorz86, Anderson87, Wen89}, and in the design of a quantum computer based on topological protection to decoherence \cite{Kitaev03, Nayak07}. Due to a lack of theoretical and computational tools, assessing whether the low energy sector of $H$ is indeed topologically ordered, and then establishing what type of topological order it realizes, have traditionally been considered very difficult problems. However, recent advances in the understanding of many-body entanglement have progressively brought us closer to being able to tackle them. On the theoretical side, new ways of diagnosing the presence and type of topological order from knowledge of the ground state wave-function alone, based on entanglement entropy \cite{Kitaev06, Levin06}, entanglement spectrum \cite{Li08} and modular transformations \cite{Zhang12, Grover11}, have been put forward. On the computational side, the advent of tensor networks \cite{White92, Verstraete04, Vidal07, Jordan08, Aguado08, Yan11, Jiang12, Depenbrock12} makes it now possible, by mimicking the structure of entanglement, to efficiently represent a large class of low energy many-body states.

In this
paper, by combining and building upon the above developments, we describe an approach for characterizing the topological order emerging from a microscopic Hamiltonian $H$ in a two dimensional lattice. The key of our approach is the computation of a basis of ground states of $H$, first on an infinite cylinder, and then on a finite torus. Each ground state has a flux through the cylinder/torus corresponding to one of the charges of the emergent anyon model. These ground states are encoded in a tensor network representation, from which we can extract the universal properties of the emergent edge and bulk theories. For concreteness, we focus on a specific lattice Hamiltonian, namely the Haldane model \cite{Haldane88} on the honeycomb lattice with hard-core bosons \cite{Wang11},
\begin{eqnarray}\label{eq:Haldane}
\HH &=& -t\sum_{\langle rr' \rangle}  b_{r}^{\dagger}b_{r'}
  ~  - t' \!\!\sum_{\langle\langle rr' \rangle\rangle} b_{r}^{\dagger}b_{r'}e^{i\phi_{rr'}} \nonumber \\
&& -t''\!\!\sum_{\langle\langle\langle rr' \rangle\rangle\rangle}  \!\!b_{r}^{\dagger}b_{r'} ~~~+~~~ ~\mbox{H.c.},
\end{eqnarray}
where we set $\phi = 0.4\pi$ and $(t,t',t'')=(1,0.6,-0.58)$, where $t,t'$ and $t''$ stand for the hoping amplitudes between nearest, next-nearest and next-next-nearest neighbors, respectively. For these parameters, Ref. \cite{Wang11} found two quasi-degenerate ground states and a non-trivial Chern number $C=1$ using exact diagonalization on small tori, while Ref. \cite{Jiang12} computed a non-zero topological entanglement entropy (TEE) \cite{Kitaev06,Levin06} $\gamma \approx \log \sqrt{2}$ by studying finite cylinders with the density matrix renormalization group (DMRG) \cite{White92}. These results are suggestive of a $\nu = 1/2$ bosonic FQH state \cite{Laughlin83}, which has a chiral semion \cite{Bonderson07}
in the bulk and a $SU(2)_1$ Wess-Zumino-Witten (WZW) conformal field theory (CFT) \cite{DiFrancesco} at the edge. Here we will not make use of any previous knowledge about $\HH$, and yet will produce a detailed characterization of both bulk and edge theories. Our approach can be readily applied to other Hamiltonians on honeycomb, triangular and kagome lattices (with $\pi/3$ rotational symmetry), and can be generalized to arbitrary lattices (even without any rotational symmetry) \cite{Lukasz12}.

\begin{figure}
  \begin{centering}
\includegraphics[width=8.0cm]{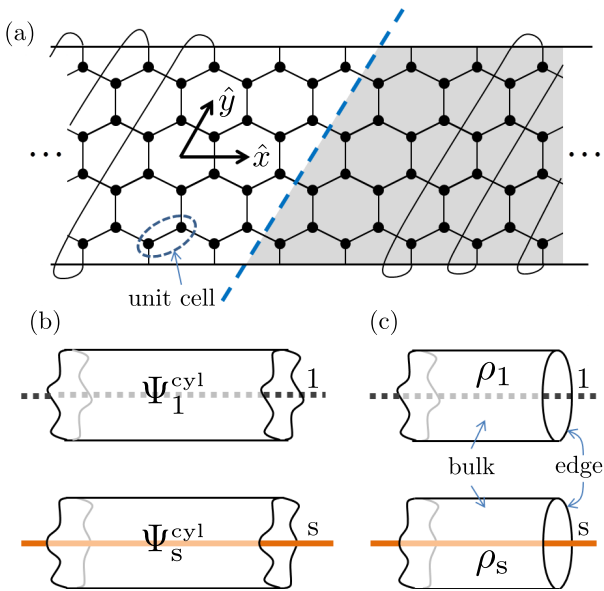}  \end{centering}
  \caption{%cylinder.eps
  (Color online) (a) Honeycomb lattice on a cylinder with $L_x=\infty$ and $L_y = 4$ (as measured in number of unit cells). A blue, dashed line indicates the entanglement cut used in dividing the cylinder into two semi-infinite cylinders. (b) Ground states $\ket{\Psi_1^{\Cyl}}$ and $\ket{\Psi_{\sss}^{\Cyl}}$ of $\HH$ in Eq. \ref{eq:Haldane} with flux $i=1,\mbox{s}$. (c) Corresponding density matrices $\rho_1$ and $\rho_{\sss}$ for the left half of the cylinder.
}
  \label{fig:cylinder}
\end{figure}

\textit{Complete set of ground states on an infinite cylinder}.--- Consider a local Hamiltonian $H$ on a cylinder of size $L_x\times L_y$, with infinite length $L_x = \infty$ and finite width $L_y$, see Fig. \ref{fig:cylinder}(a). If the width $L_y$ was also infinite, $H$ would have a finite number $N$ of exactly degenerate ground states, one for each species $i$ of anyon in the emergent anyon model \cite{Wen90}. A finite $L_y$ implies, however, that the ground state degeneracy is broken. This is due to the creation of virtual pairs of anyons that annihilate after describing a non-trivial loop in the $\hat{y}$ direction \cite{Kitaev03}. If the width $L_y$ of the cylinder is much larger than the correlation length $\xi$ in the ground states of $H$, $L_y \gg \xi$, we expect the topological order to be still (approximately) realized, and the "low energy" space of $H$ to still decompose into $N$ distinct topological sectors. Each of these sectors is characterized by a ground state $\ket{\Psi_i^{\Cyl}}$,
\begin{equation}\label{eq:Psii}
    H \ket{\Psi_i^{\Cyl}} = E_i \ket{\Psi_i^{\Cyl}},
\end{equation}
which, \textit{crucially for the present approach}, has an anyon of type $i$ propagating inside the cylinder in the $\hat{x}$ direction, see Fig. \ref{fig:cylinder}(b) -- that is, $\ket{\Psi_i^{\Cyl}}$ is an eigenvector of the \textit{dressed} Wilson loop operator \cite{Hastings05} encircling the cylinder in the $\hat{y}$ direction. This is so because virtual pairs of anyons encircling the cylinder in the $\hat{x}$ direction only renormalize the energy $E_i$ of $\ket{\Psi^{\Cyl}_i}$
(see also Appendix A).
[It can be further argued, using degenerate perturbation theory, that the gap $\Delta_{ij} \equiv E_i - E_j$ between ground state energies is exponentially suppressed with $L_y$  but extensive in $L_x$, $\Delta_{ij} \approx e^{-aL_y}L_x$, and thus infinite. An infinite gap $\Delta_{ij}$ does not jeopardize the present approach.]

\begin{figure}
  \begin{centering}
\includegraphics[width=9.0cm]{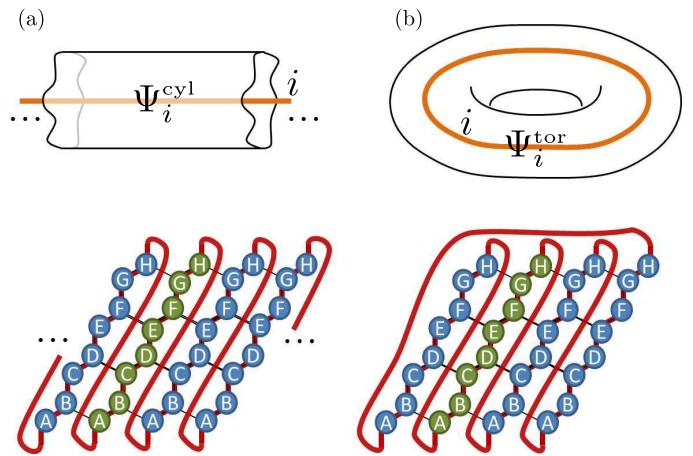}  \end{centering}
  \caption{%mps.eps
  (Color online)(a) MPS for a state $\ket{\Psi_i^{\Cyl}}$ of an infinite cylinder with $L_y=4$, see Fig. \ref{fig:cylinder}(a). The bond indices that connect tensors form a snake. The MPS unit cell is made of eight tensors that are repeated throughout the tensor network. (b) MPS for a state $\ket{\Psi_i^{\Tor}}$ of a torus with $L_x \times L_y \times 2= 32$ sites, see Fig. \ref{fig:modular}(a), obtained by reconnecting four MPS unit cells of the state $\ket{\Psi_i^{\Cyl}}$.
}
  \label{fig:mps}
\end{figure}

\textit{Tensor networks.}---
Each ground state $\ket{\Psi^{\Cyl}_i}$ can be represented using a tensor network with a finite unit cell of tensors that is repeated throughout the infinite cylinder
(see Appendix B).
A natural choice is to use a projected entangled-pair state (PEPS) \cite{Verstraete04}, optimized e.g. through energy minimization \cite{Verstraete04} or imaginary time evolution \cite{Jordan08}. The incurred computational cost is at most linear in $L_y$, which allows for the constraint $L_y \gg \xi$ to be easily met. In the present application we study the ground states of $H_{\mbox{\tiny Hald.}}$ using instead a matrix product state (MPS) optimized using DMRG \cite{White92}, see Fig. \ref{fig:mps}(a).
The computational cost grows now exponentially with the width $L_y$ of the cylinder, and therefore only small values of $L_y$ can be afforded. However, if the condition $L_y \gg \xi$ can still be met, DMRG offers a more accurate and dependable implementation of our approach.

Specifically, with an infinite DMRG \cite{McCulloch08, Crosswhite08} algorithm in two dimensions (detailed in Ref. \cite{Lukasz12}), we considered infinite cylinders with width $L_y=4,6$ and $8$ (as measured in number of unit cells, see Fig. \ref{fig:cylinder}). For each value of $L_y$, we repeated the optimization of an infinite MPS several times. Each optimization produced one of two states, with each state occurring roughly half of the runs. We thus conclude that there are at least two ground states or, equivalently, that the emergent anyon model has $N\geq 2$ anyon charges. Let us denote these two ground states $\ket{\PsiId^{\Cyl}}$ and $\ket{\PsiSe^{\Cyl}}$, anticipating that they correspond to anyon types $i=\mbox{1}$, or \textit{identity} charge, and $i=\mbox{s}$, or \textit{semion} charge, respectively, to be confirmed later. A very small correlation length $\xi_{1} \approx \xi_{\sss} \approx 0.52$ for both states verifies that $L_y \gg \xi$ and thus we have effectively reached the thermodynamic limit.

\begin{figure}
  \begin{centering}
\includegraphics[width=8.5cm]{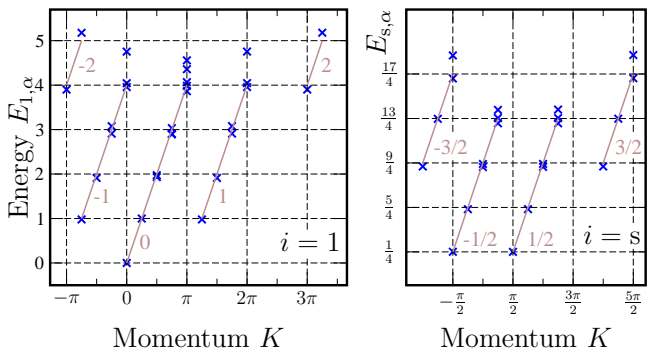}  \end{centering}
  \caption{%CFTtowers.eps
  (Color online)
  Entanglement spectrum of the reduced density matrix $\rho_i$ for half an infinite cylinder for ground state $\ket{\Psi_i^{\Cyl}}$, for $i=1$ and $\mbox{s}$, Fig. \ref{fig:cylinder}(a). The vertical axis shows $E_{i,\alpha} \equiv -\log (p_{i,\alpha})$ (up to global shift and re-scaling), where $\{p_{i,\alpha}\}$ are the eigenvalues of $\rho_i$. The horizontal axis shows the momentum of the corresponding eigenvector of $\rho_i$, which we artificially extend beyond its $2\pi$ periodicity. A tilted line connects all eigenvalues of $\rho_i$ with the same particle number $M$, also indicated (notice that $\HH$ has a $U(1)$ symmetry corresponding to particle number conservation). We find a degeneracy pattern $\{1,1,2,3,5, \cdots\}$ (according to the partition numbers) along each tilted line of fixed $M$, characteristic of a bosonic Gaussian theory.
}
  \label{fig:CFTtowers}
\end{figure}

\textit{Quantum dimensions from entanglement entropy}.---
The tensor network representation of state $\ket{\Psi_i^{\Cyl}}$ readily provides access to the eigenvalues $\{p_{i,\alpha}\}$ of the reduced density matrix $\rho_i$ for half of the infinite cylinder, see Fig. \ref{fig:cylinder}(c). From the scaling of the entanglement entropy $S(\rho_i)\equiv - \sum_{\alpha} p_{i\alpha} \log(p_{i\alpha})$ as a function of the width  $L_y$, namely $S_i({L_y}) = \alpha L_y - \gamma_i$, we can estimate the TEE $\gamma_i$ \cite{Kitaev06, Levin06} for ground state $\ket{\Psi_i^{\Cyl}}$. We obtain $\gamma_1 \approx 0.3455$, $\gamma_{\sss} \approx 0.986 \times \gamma_1$, very similar to the finite cylinder result of Ref. \cite{Jiang12} for one ground state.
The expression \cite{Kitaev06,Dong08}
\begin{equation}\label{eq:quantumD}
\gamma_i = -\log (d_i/D),~~~~~~~~~~~D \equiv \sqrt{\sum_i (d_i)^2},
\end{equation}
allows us to compute each \textit{quantum dimension} $d_i$ and the \textit{total quantum dimension} $D$
(see Appendix D).
First, we note that the sum $\sum_i (d_i/D)^2$ only reaches unity when it includes all the anyon charges. We can therefore use this sum to establish whether we have obtained a complete set of ground states. Our estimates of $\gamma_i$ yield $(d_1/D)^{2}+(d_{\sss}/D)^2 \approx 1.007$. Thus, $\HH$ only has two ground states, \textit{i.e.} the emergent anyon model has $N=2$ types of anyons \cite{Wang11}. Then, since the identity charge always exists and has quantum dimension $d_1=1$, from $\gamma_{\sss}-\gamma_1=-\log(d_{\sss}/d_1)$ we find $d_{\sss} = 1.005$, and from $\gamma_1=-\log(d_1/D)$ we find $D = 1.413$, very close to the exact $d_{\sss} = 1$ and $D = \sqrt{2} \approx 1.4142$ of the semion model \cite{Bonderson07}.

\textit{Edge chiral CFT from entanglement spectrum}.---
The entanglement spectrum (ES) \cite{Li08} for $\ket{\Psi_i^{\Cyl}}$
(see Appendix C),
plotted in Fig. \ref{fig:CFTtowers}, reveals that each ground state of $\HH$ is associated with one of the two primary field of the chiral $SU(2)_1$ WZW CFT, namely the identity and chiral boson vertex operator $e^{i \phi/\sqrt{2}}$ (see e.g. Chap 15.6 of Ref. \cite{DiFrancesco}), which are seen to be a singlet and a doublet of an emergent $SU(2)$ symmetry. Specifically, the ES of $\ket{\PsiId^{\Cyl}}$ is organized according to the scaling dimensions and conformal spins of the identity primary field of this CFT and the tower of all of its (Kac-Moody and Virasoro) descendants, whereas the ES of $\ket{\PsiSe^{\Cyl}}$ corresponds to the chiral boson vertex operator $e^{i \phi/\sqrt{2}}$ and its descendants, which justifies our previous identification of $\ket{\Psi_1^{\Cyl}}$ and $\ket{\Psi_{\sss}^{\Cyl}}$ with the anyonic charges $i=1$ and $\mbox{s}$, respectively. Thus, from the states $\{\ket{\Psi_i^{\Cyl}}\}$ we have unambiguously identified the edge CFT of the emergent anyon model.

\textit{From an infinite cylinder to a finite torus}.--- Using an \textit{infinite} cylinder has the major advantage, compared to previous DMRG studies on a \textit{finite} cylinder \cite{Yan11, Jiang12, Depenbrock12}, that it provides a complete set of ground states $\{\ket{\Psi_i^{\Cyl}}\}$ of a local Hamiltonian $H$. In addition, we can also produce a complete basis $\{\ket{\Psi_i^{\Tor}}\}$ for the (quasi-degenerate) ground space of $H$ on a finite torus of size $L_x\times L_y$, where the choice $L_x = L_y$ ensures that also $L_x \gg \xi$. This is accomplished by reconnecting a region of size $L_x \times L_y$ of the tensor network for $\{\ket{\Psi^{\Cyl}_i}\}$ into a torus, see Fig. \ref{fig:mps}(b)
(and Appendix B).
Notice that state $\ket{\Psi^{\Tor}_i}$ has an anyon of type $i$ inside the torus. Following Ref. \cite{Zhang12} we can use the basis $\{\ket{\Psi^{\Tor}_i}\}$ to characterize the emergent anyon theory of $H$ \cite{Note}.

\textit{Bulk anyon model from modular transformations}.---
Our goal is to compute the modular matrices $S$ and $U$ characterizing the mutual and self statistics of the emergent anyon model,
\begin{equation}\label{eq:SU}
S =
\left[
  \begin{array}{cc}
    S_{11} & S_{1\sss} \\
    S_{\sss 1} & S_{\sss\sss} \\
  \end{array}
\right],~~~~
U = e^{-i\frac{2\pi}{24}c}
\left[
  \begin{array}{cc}
    \theta_{1} & 0 \\
    0 & \theta_{\sss} \\
  \end{array}
\right],
\end{equation}
Here,
entry $S_{ij}$ determines the phase acquired by an anyonic charge $i$ when encircling an anyonic charge $j$; $c$ is the central charge of the anyon model; and $\theta_i$ corresponds to the phase acquired when an anyonic charge $i$ is exchanged with another identical anyonic charge $i$ (and thus $\theta_{1} = 1$). From  $S_{1i}=S_{i1}=d_i/D$ and $\sum_i (S_{1i})^2 = 1$ ($S$ is a unitary matrix) we can also extract the quantum dimensions $d_i$ and total quantum dimension $D$, whereas the fusion rules $i \times j = \sum _k {N_{ij}}^k ~ k$ follow from the Verlinde formula ${N_{ij}}^k = \sum_m (S_{im}S_{jm}(S_{mk})^{*}/S_{1m})$ \cite{Verlinde}.

\begin{figure}
  \begin{centering}
\includegraphics[width=8.5cm]{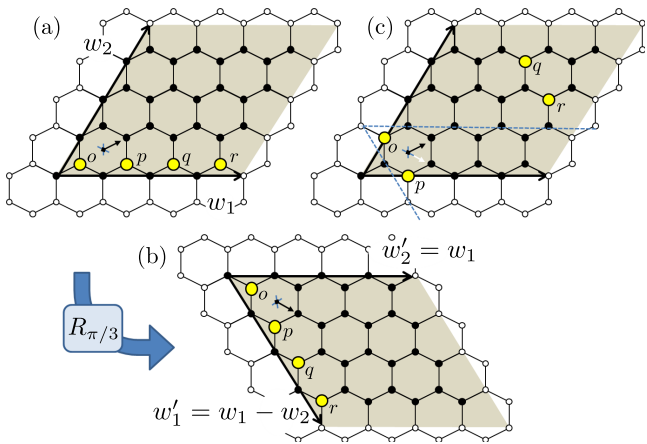}  \end{centering}
  \caption{
  %modular.eps
  (Color online) (a) Torus with $L_x=L_y=4$, made of $L_x\times L_y \times 2 = 32$ sites, and with generating vectors $w_1$ and $w_2$. Four sites are marked as $o,p,q,r$ for reference. (b) Torus obtained after a $\pi/3$ rotation $R_{\pi/3}$, corresponding to a modular transformation that maps $(w_1,w_2)$ into $(w_1',w_2') = (w_1-w_2, w_1)$. (c) By using the periodicity by vectors $(w_1,w_2)$ [equivalently, ($w_1', w_2'$)], the sites in torus (b) can be mapped back into torus (a). This allows us to compare states defined on the two tori and compute $V$ in Eq. \ref{eq:V}, following Ref. \cite{Zhang12}.
}
  \label{fig:modular}
\end{figure}

 A $\pi/3$ rotation $R_{\pi/3}$ on the torus, see Fig. \ref{fig:modular}, corresponds to applying the modular transformation $\mathfrak{u} \mathfrak{s}^{-1}$, where  $\mathfrak{u}$ and $\mathfrak{s}$ generate the group of modular transformations on a torus with defining vectors $w_1$ and $w_2$,
\begin{equation}\label{eq:sANDu}
    \mathfrak{s}:\left[
         \begin{array}{c}
           w_1 \\ w_2 \\
         \end{array}
       \right] \mapsto
       \left[
         \begin{array}{c}
           w_2 \\ -w_1 \\
         \end{array}
       \right],~~
    \mathfrak{u}: \left[
         \begin{array}{c}
           w_1 \\ w_2 \\
         \end{array}
       \right] \mapsto
       \left[
         \begin{array}{c}
           \!w_1 \! + \! w_2 \!\\ w_2 \\
         \end{array}
       \right]. \nonumber
\end{equation}
 Following Ref. \cite{Zhang12}, the overlaps $V_{ij} \equiv \bra{\Psi_i^{\Tor}} R_{\pi/3} \ket{\Psi_j^{\Tor}}$ between the bases $\{\ket{\Psi_i^{\Tor}}\}$ and $\{R_{\pi/3} \ket{\Psi_j^{\Tor}}\}$ form the unitary transformation $V = D^{\dagger} U S^{-1} D$, where $D$ is a diagonal matrix of phases $D_{jj} = e^{i\phi_j}$ corresponding to a phase freedom in choosing $\ket{\Psi_j^{\Tor}}$, and $S$ and $U$ generate a representation of the modular group on the ground space of $\HH$. We thus have
\begin{eqnarray}
    V &=& e^{-i\frac{2\pi}{24}c}
    \left[
      \begin{array}{ll}
(S_{11})^{*}\theta_1 & (S_{\sss 1})^{*} \theta_1 e^{i(\phi_{\sss} - \phi_1)}\\
(S_{1\sss})^{*}\theta_{\sss}e^{i(\phi_1 - \phi_{\sss})} &(S_{\sss\sss})^{*} \theta_{\sss}\\
      \end{array}
    \right] \nonumber \\
    &=& e^{-i\frac{2\pi}{24}c}
    \left[
      \begin{array}{ll}
S_{11} & S_{\sss 1} e^{i(\phi_{\sss} - \phi_1)}\\
S_{1\sss} \theta_{\sss} e^{i(\phi_1 - \phi_{\sss})} &(S_{\sss\sss})^{*} \theta_{\sss}\\
      \end{array}
    \right],\label{eq:V}
\end{eqnarray}
where in the second line we use that $S_{1i},S_{i1}>0$ and $\theta_{1}=1$. Building upon Ref. \cite{Zhang12}, we note that from matrix $V$ we can actually compute \textit{both} $U$ and $S$. Indeed, from $V_{11}$ we obtain the central charge $c$ and $S_{11}$; then from $V_{1\sss}$ we obtain $e^{i(\phi_{\sss} - \phi_1)}$ and $S_{\sss 1}$; then from $V_{\sss 1}$ we obtain $\theta_{\sss}$ and $S_{1 \sss}$; finally, $V_{\sss \sss}$ yields $S_{\sss \sss}$. [Similarly, for an arbitrary anyon model, from the scalar products $V_{ij}$ we can completely determine both $S_{ij}$ and $U_{ij}$,
see Appendix E].
The \textit{exact} evaluation of $V_{ij}$ from a tensor network representation turns out to have a very large computational cost. We compute $V_{ij}$ \textit{approximately} using Monte Carlo sampling on a periodic MPS \cite{Sandvik07}. For $L_y=6$ (torus with $6 \times 6 \times 2 = 72$ sites), we obtain

\begin{eqnarray}\label{eq:numS}
    S &=& \frac{1}{\sqrt{2}}\left[ \begin{array}{cc} 1 & 1 \\1 & -1 \\\end{array} \right]
    + \frac{10^{-3}}{\sqrt{2}}\left[
          \begin{array}{cc}
            -1.4 & 0.2 \\
            -1.4 & 4+4i \\
          \end{array}
        \right], \\ \label{eq:numU}
    U &=& e^{-i\frac{2\pi}{24}}\left[
          \begin{array}{cc}
            1 & 0 \\
            0 & i \\
          \end{array}
        \right]\times \left(e^{i\frac{2\pi}{24}0.01}\left[
          \begin{array}{cc}
            1 & 0 \\
            0 & e^{-i 0.007} \\
          \end{array}
        \right] \right),
\end{eqnarray}
 with sampling noise on the order of $10^{-3}$, which accurately reproduce the exact $S$ and $U$ matrices of a chiral semion model, namely $\frac{1}{\sqrt{2}}\left[ \begin{array}{cc} 1 & 1 \\1 & -1 \\\end{array} \right]$,
and $e^{-i\frac{2\pi}{24}} \left[ \begin{array}{cc} 1 & 0 \\ 0 & i \\ \end{array} \right]$. In particular, from Eq. \ref{eq:numS} we can accurately extract the quantum dimensions $d_1=d_{\sss}=1$ and total quantum dimension $D=\sqrt{2}$ (independently of the previous entropy calculations) and the $\mathbb{Z}_2$ fusion rules $1\times 1 = \mbox{s} \times \mbox{s} = 1$ and $1\times \mbox{s} = \mbox{s} \times 1 = \mbox{s}$; whereas from Eq. \ref{eq:numU} we determine a central charge $c = 1$ (modulus 24) and the twist $\theta_{\sss} = i$ characteristic of a semion.

In this
paper
we have shown how to produce, starting from a microscopic lattice Hamiltonian $H$, a detailed characterization of the emergent topological order. A key step is to obtain a tensor network representation for a complete set of ground states of $H$, first on an infinite cylinder and then on a finite torus. Here we have considered several ground state properties that can be extracted directly from the optimized tensor network. With further manipulation, it is also possible to obtain an explicit representation of quasi-particle excitations of $H$, including those with fractional quantum numbers, as discussed elsewhere \cite{Lukasz12}.

%%%%%%%%%%%%%%%%%%%%%%%

The authors thank Jaume Gomis and Xiao-Gang Wen for guidance in identifying the edge theory resulting from the numerical simulations, and Oliver Buerschaper, Tarun Grover, Todadri Senthil, and Ashvin Vishwanath for insightful discussions.

%\newpage

\appendix

\section{Ground states on an infinite cylinder}

Let $H$ be a local lattice Hamiltonian corresponding to a topologically ordered phase, and let $\xi$ be the largest correlation length in the ground space of $H$ on an infinite lattice. In this appendix we argue that, when placed on a torus of size $L_x \times L_y$ with $L_x = \infty$ and $L_y \gg \xi$, each of the (quasi-degenerate) ground states $\{\ket{\Psi_i}\}$ of $H$ has a well-defined anyon flux $i$ in the $\hat{x}$ direction, as measured by a dressed \cite{Hastings05} Wilson loop operator $\Gamma_{\hat{y}}$ encircling the torus in the $\hat{y}$ direction, see Fig. \ref{fig:torus}. Then we argue that when placed on a cylinder of size $L_x \times L_y$ with $L_x = \infty$ and $L_y \gg \xi$ (and convenient boundary conditions to be specified below), $H$ also has analogous (quasi-degenerate) ground states $\{\ket{\Psi^{\Cyl}_i}\}$. This last set of ground states can be represented and computed with tensor network techniques, as discussed in the main text of this paper and in Appendix B.

\subsection{Ground states of a fixed-point Hamiltonian $H^{(0)}$}

Let us first consider a fixed-point Hamiltonian $H^{(0)}$ describing the same phase as $H$ (but with vanishing correlation length, $\xi = 0$). We will assume that $V \equiv H-H^{(0)}$ is a small perturbation that can be expressed as a sum of local terms and treated using (degenerate) perturbation theory.

\begin{figure}
  \begin{centering}
\includegraphics[width=8.0cm]{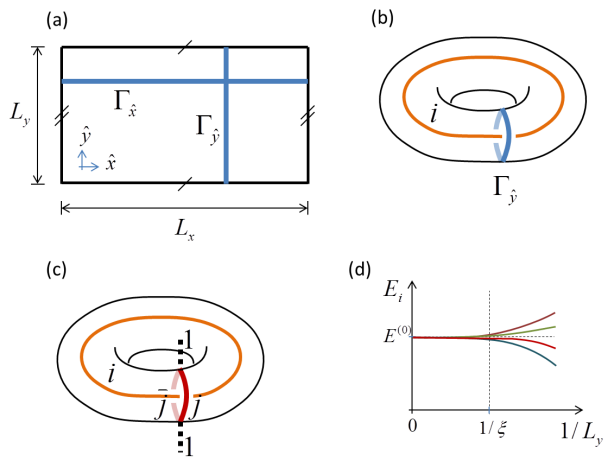}  \end{centering}
  \caption{%torus.eps
  (Color online) (a) Torus of size $L_x \times L_y$. A Wilson loop operator $\Gamma_{\hat{y}}$ encercling the torus in the $\hat{y}$ direction measures the anyon flux in the $\hat{x}$ direction. (b) The same torus, with anyonic flux $i$ in the $\hat{x}$ direction. (c) Modifying a fixed-point Hamiltonian $H^{(0)}$ with a perturbation $V$ has the effect, in perturbation theory, of creating virtual pairs of anyons that propagate through the torus. The figure represents a pair of anyons $j$ and $\bar{j}$ that annihilate after closing a non-trivial loop in the $\hat{y}$ direction. This may shift the energy of $\ket{\Psi_i}$ by an amount that depends on the anyon type $i$. (d) For $L_x = \infty$ and finite $L_y$, the ground state degeneracy is broken in the $\ket{\Psi_i}$ basis.
}
  \label{fig:torus}
\end{figure}

Let $\{\ket{\Psi_i^{(0)}}\}$ denote a basis of ground states of $H^{(0)}$ on a torus of size $L_x \times L_y$, where state $\ket{\Psi_i^{(0)}}$ has anyon flux $i$ in the $\hat{x}$ direction, i.e., $\ket{\Psi_i^{(0)}}$ is an eigenvector of the Wilson loop operator $\Gamma^{(0)}_{\hat{y}}$ encircling the torus in the $\hat{y}$ direction, see Fig. \ref{fig:torus}(a)-(b). Clearly, the states $\{\ket{\Psi_i^{(0)}}\}$ are just one possible basis in the ground space of $H^{(0)}$, but this particular choice of basis plays a distinguished role below. We remind the reader that all ground states of the topologically ordered Hamiltonian $H^{(0)}$ are locally indistinguishable -- their reduced density matrices are identical on any contractible region -- and that in order to transform the ground state $\ket{\Psi^{(0)}_i}$ into another ground state $\ket{\Psi^{(0)}_j}$, a non-local operator closing a non-trivial loop in the $\hat{x}$ direction, such as a Wilson loop operator $\Gamma^{(0)}_{\hat{x}}$, is required.

Let us now assume that, starting with the lattice in the ground state $\ket{\Psi^{(0)}_i}$ of $H^{(0)}$, we adiabatically switch on perturbation $V$, so that Hamiltonian is changed from $H^{(0)}$ to $H$. Let $\ket{\Psi_i}$ denote the resulting state of the system,
\begin{equation}\label{eq:adiabatic}
    \ket{\Psi_i^{(0)}} \stackrel{\mbox{ \tiny adiabatic}}{\longrightarrow} \ket{\Psi_i}.
\end{equation}
For a generic choice of sizes $L_x$ and $L_y$, $\ket{\Psi_i}$ will no longer be an eigenvector of $H$. However, we will argue below that for $L_x = \infty$ and finite $L_y \gg \xi$, $\ket{\Psi_i}$ is still an eigenvector of $H$. In preparation, we first consider a torus with $(L_x,L_y)=(\infty,\infty)$.

\subsection{Ground states of $H$ on a torus with $L_x = L_y = \infty$}

For $L_x = L_y = \infty$, let us analyze the effect of perturbation $V$ (which by hypothesis does not break the topological order) using perturbation theory. Perturbation $V$ can be interpreted as to allow for the creation of virtual pairs of anyons that propagate throughout the torus \cite{Kitaev03}. If one virtual pair annihilates after closing a non-trivial loop, this may change the topological properties of the ground state. However, we notice that, due to the infinite size of the torus in any direction, no virtual pair of anyons can close a non-trivial loop at $m^{\mbox{\tiny th}}$ order in perturbation theory, for any finite $m$, because the $m^{\mbox{\tiny th}}$ power of $V$ decomposes as the sum of terms each of which still has support on a contractible region only. In other words, for any finite $m$, the overlap $\bra{\Psi^{(0)}_i} (V)^{m} \ket{\Psi^{(0)}_i}$ is equal to some value $\langle (V)^{m} \rangle$ independent of the flux $i$ (since all ground states of $H^{(0)}$ are identical on any contractible region), whereas for $i\neq j$ the overlap $\bra{\Psi^{(0)}_i} (V)^{m} \ket{\Psi^{(0)}_j}$ vanishes (because $(V)^{m}$ cannot close a non-trivial loop in the $\hat{x}$ direction),
\begin{equation}\label{eq:perturbation1}
    \bra{\Psi^{(0)}_i} (V)^{m} \ket{\Psi^{(0)}_j} = \delta_{i,j}  \langle (V)^{m} \rangle.
\end{equation}

As a result, in degenerate perturbation theory each state $\ket{\Psi_i}$ is obtained from the corresponding $\ket{\Psi_i^{(0)}}$ by means of the same local modifications, so that all the states $\ket{\Psi_i}$ remain locally indistinguishable, and form a complete set of (exactly degenerate) ground states of $H$ with renormalized energy $E$, i.e. $H \ket{\Psi_i} = E \ket{\Psi_i}$, and finite correlation length $\xi$. In particular, state $\ket{\Psi_i}$ still has anyon flux $i$ in the $\hat{x}$ direction, as measured now by a \textit{dressed} \cite{Hastings05} Wilson loop operator $\Gamma_{\hat{y}}$ in the $\hat{y}$ direction.

\subsection{Ground states of $H$ on a torus with $L_x = \infty$ and finite $L_y \gg \xi$}

For $L_x = \infty$ and finite $L_y \gg \xi$, the ground state degeneracy of $H$ is broken in perturbation theory due to the fact that virtual pairs of anyons can now annihilate with each other after their combined trajectories have closed a non-trivial loop in the $\hat{y}$ direction, see Fig. \ref{fig:torus}(c), producing an energy contribution $\langle (V)^{m} \rangle_i \equiv \bra{\Psi^{(0)}_i} (V)^{m} \ket{\Psi^{(0)}_i}$ (occurring first for some finite $m$ proportional to $L_y$) that may now depend on the flux $i$ present in the $\hat{x}$ direction of the cylinder. On the other hand, no virtual loop can still be closed in the $\hat{x}$ direction for any finite $m$, since $L_x = \infty$, and thus $\bra{\Psi^{(0)}_i} (V)^{m} \ket{\Psi^{(0)}_j} = 0$ for $i\neq j$, or
\begin{equation}\label{eq:perturbation2}
    \bra{\Psi^{(0)}_i} (V)^{m} \ket{\Psi^{(0)}_j} = \delta_{i,j} \langle (V)^{m} \rangle_i.
\end{equation}

As a result, the ground state degeneracy is broken precisely in the $\{\ket{\Psi_i}\}$ basis, so that each $\ket{\Psi_i}$ remains an exact eigenvector of $H$ with (now distinct) energy $E_i$, $H\ket{\Psi_i} = E_i \ket{\Psi_i}$. A gap $\Delta_{ij} \equiv E_i - E_j$ opens. Since this first occurs in perturbation theory at a finite order $m$ that is proportional to $L_y$, the gap decays exponentially in $L_y$, $\Delta_{ij} \approx e^{-a L_y}$, for some constant $a$ \cite{Kitaev03}. However, tunneling of virtual anyon pairs in the $\hat{y}$ direction can occur anywhere in the $\hat{x}$ direction, and therefore the gap is also proportional to $L_x$, $\Delta_{ij} \approx e^{-a L_y}L_x$, and thus infinite. [This will turn out to be of no relevance in our numerical simualtions based on energy minimization, because the simulation still gets trapped in some topological sector, independently of the gaps $\Delta_{ij}$]. Importantly, once again the state $\ket{\Psi_i}$ has anyon flux $i$ in the $\hat{x}$ direction, as measured now by a \textit{dressed} \cite{Hastings05} Wilson loop operator $\Gamma_{\hat{y}}$ in the $\hat{y}$ direction. We continue to refer to state $\ket{\Psi_i}$ as a ground state of $H$, since it is its ground state in the $i$ flux sector.

\begin{figure}
  \begin{centering}
\includegraphics[width=8.0cm]{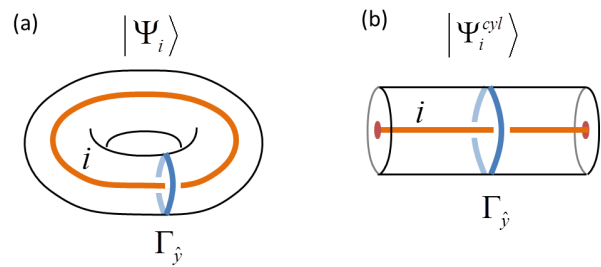}  \end{centering}
  \caption{%torus2.eps
  (Color online) (a) Schematic representation of a ground state $\ket{\Psi_i}$ of $H$ on a torus of size $L_x\times L_y$ with $L_x = \infty$ and finite $L_y$, with anyon flux $i$ in the $\hat{x}$ direction. (b) Schematic representation of an analogous ground state $\ket{\Psi_i^{\Cyl}}$ of $H$ on a cylinder of the same size $L_x \times L_y$, where the anyon flux is created and annihilated at the $x = \pm \infty$ boundaries.
 }
  \label{fig:torus2}
\end{figure}

\subsection{Local equivalence between the ground states of $H$ on a torus and a cylinder when $L_x = \infty$ and finite $L_y \gg \xi$.}

We have argued above that the combined effect of perturbation $V$ and finite $L_y\ll \xi$ on the torus is that Hamiltonian $H$ has eigenstates $\ket{\Psi_i}$ (referred to as ground states) with well defined anyon flux $i$ in the $\hat{x}$ direction.

Let us now consider a cylinder of size $L_x \times L_y$, with $L_x = \infty$ and finite $L_y \gg \xi$, and a Hamiltonian on this cylinder that decomposes into: (i) a bulk Hamiltonian that is locally identical to the Hamiltonian $H$ above for the torus, and (ii) boundary Hamiltonian terms $H^{-\infty}$ and $H^{+\infty}$ for the $x = -\infty$ and $x=\infty$ boundaries of the cylinder. [Here $L_x = \infty$ describes the limit of a cylinder with finite $L_x$, for which an analogous Hamiltonian can be defined]. We assume that $H^{\pm \infty}$ is such that an anyon of type $i$ can be created at one boundary and annihilated at the other boundary at a finite energy cost. Then the ground states of $H^{\Cyl}$, denoted \{$\ket{\Psi_i^{\Cyl}}$\}, have anyon flux $i$ through the cylinder, see Fig. \ref{fig:torus2}, as can be argued in a similar way as we discussed above for a torus. In addition, on any finite section of the cylinder $\ket{\Psi_i^{\Cyl}}$ is identical to the analogous ground state $\ket{\Psi_i}$ on the torus, since these states only have short correlations and result from minimizing a local Hamiltonian, which is the same on the torus than in the bulk of the cylinder.

The particular choice of the above boundary Hamiltonian terms $H^{\pm \infty}$ is actually of no practical importance for our numerical approach. Indeed, in a tensor network computation of the ground states of an infinite cylinder, as used in this paper, the optimization of the bulk tensors will not take into account energy contributions coming from the boundary.

\begin{figure}
  \begin{centering}
\includegraphics[width=8.0cm]{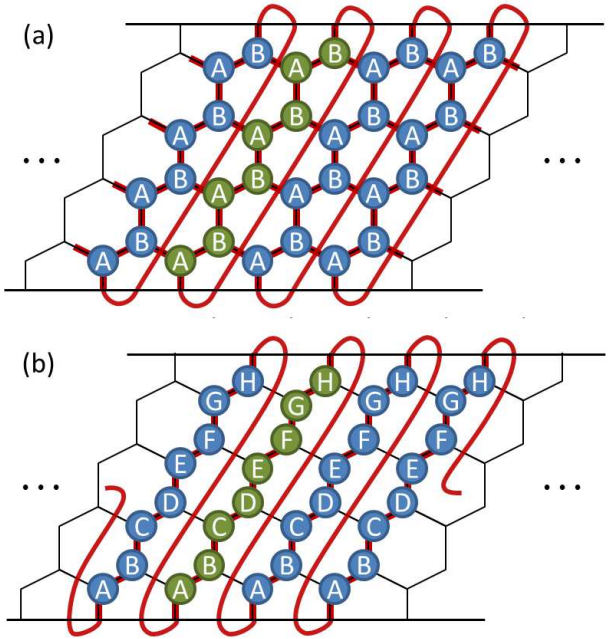}  \end{centering}
  \caption{%PepsMpsCyl.eps
  (Color online) (a) PEPS representation for a state $\ket{\Psi^{\Cyl}}$ on a cylinder of size $(L_x,L_y) = (\infty, 4)$ (honeycomb lattice). Notice that the PEPS consists in repeating two tensors $A$ and $B$, and thus the PEPS unit cell coincides with the natural unit cell on the honeycomb lattice. (b) MPS representation for the same state. Here the MPS unit cell is made of eight different tensors $\{A, B, \cdots, H \}$, connected in a one-dimensional geometry that snakes around the cylinder. Each MPS unit cell corresponds to four (in general, $L_y$) unit cells of the honeycomb lattice.
  }
  \label{fig:PepsMpsCyl}
\end{figure}

\section{Tensor network representations on an infinite cylinder}

In this work we have used a tensor network to represent states of a cylinder of size $L_x \times L_y$, where $L_x=\infty$ and $L_y$ is finite. We consider two different tensor networks: the projected entangled-pair state (PEPS) and the matrix product states (MPS).

\subsection{PEPS on an infinite cylinder}

A natural option is to use a projected entangled pair state (PEPS) \cite{Verstraete04} with periodic boundary conditions in the $\hat{y}$ directions. In the presence of translation invariance, the state of an infinite system can be represented by repeating a unit cell of tensors throughout the lattice \cite{Jordan08}. For instance, in a honeycomb lattice, two tensors  $A$ and $B$ can be used to form a PEPS unit cell, see Fig. \ref{fig:PepsMpsCyl}(a). In this example, the PEPS unit cell coincides with the natural unit cell on the honeycomb lattice. Notice that this representation is very compact. Tensors $A$ and $B$ have components $(A)_{\alpha \beta \gamma s}$ and $(B)_{\alpha \beta \gamma s}$, where the \textit{bond indices} $\alpha,\beta$, and $\gamma$ take $D$ values each (e.g. $\alpha = 1, \cdots,D)$ and the \textit{physical index} takes $d$ values, where $d$ is the dimension of the Hilbert space of one lattice site ($d=2$ in Hamiltonian $\HH$, since on each site we may have zero or one bosons). This amounts to about $2dD^3$ complex coefficients, or variational parameters, to characterize the PEPS. For a gapped phase, the value of $D$ necessary to obtain an accurate description of a ground state $\ket{\Psi_i^{\Cyl}}$ is not expected to depend on $L_y$.  Importantly, the computational cost of optimizing these tensors can also be made essentially independent of the size $L_y$ of the cylinder, which allows us to consider large values of $L_y$.

\subsection{MPS on an infinite cylinder}

An alternative tensor network representation for states of an infinite cylinder is a matrix product state (MPS), which is the basis of the density matrix renormalization group (DMRG) \cite{White92}. In the example of Fig. \ref{fig:PepsMpsCyl}(b), a unit cell of tensors is also repeated throughout the cylinder, but the MPS unit cell is $L_y$ times the size of the natural unit cell on the honeycomb lattice. An MPS tensor $A$ has components $(A)_{\alpha\beta s}$, where the bond indices $\alpha$ and $\beta$ take $\chi$ values and the physical index $s$ takes $d$ values. Thus, the number of variational parameters in an MPS for an infinite cylinder is of the order of $2L_yd\chi^2$. Key to this discussion is the fact that the value of $\chi$ necessary to obtain an accurate description of a ground state $\ket{\Psi_i^{\Cyl}}$ is expected to grow exponentially with $L_y$, making the MPS representation (and its manipulation, with cost scaling as $\chi^3$) inefficient. As a result, only systems with small values of $L_y$ can be afforded computationally. For states with a sufficiently small correlation length $\xi$ (as given, for instance, by the transfer matrix correlation length $\xi_{\mbox{\tiny TM}}$, see next subsection), such small values of $L_y$ might already be sufficient in order to make finite-size effects negligible and thus obtain the thermodynamic properties of the system. This is precisely the case with the ground states of Hamiltonian $\HH$ in Eq. \ref{eq:Haldane}.

Twenty years of successful DMRG experience makes this approach much better understood than PEPS, which is only a few years old. DMRG offers a more accurate and dependable implementation of our approach whenever it can be applied -- that is, when the condition $L_y \gg \xi$ can be met for sufficiently small cylinder width $L_y$.

\begin{figure}
  \begin{centering}
\includegraphics[width=8.0cm]{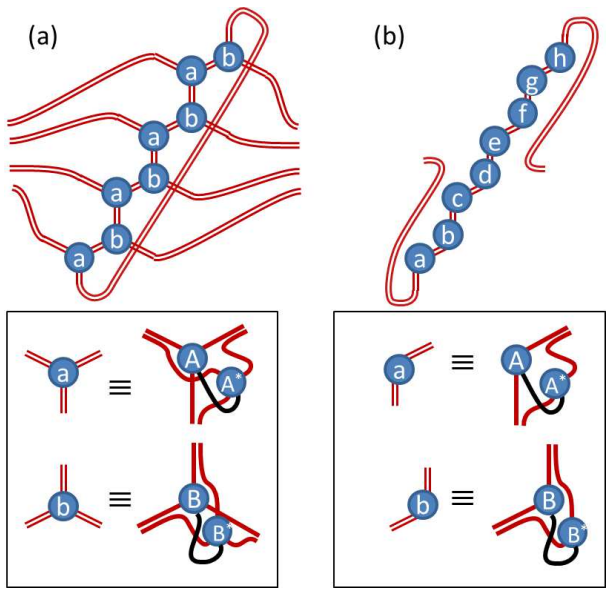}  \end{centering}
  \caption{%PepsMpsTM.eps
(Color online) (a) The PEPS transfer matrix on the $\hat{x}$ direction is obtained by building the reduced tensors $a$ and $b$, see Eq. \ref{eq:a}, and connecting these tensors periodically in the $\hat{y}$ direction. The inset shows tensors $a$ and $b$ in terms of tensors $A$, $A^{*}$, and $B$, $B^{*}$, respectively. (b) The MPS transfer matrix is built in an analogous way. The inset shows two of the reduced tensors, $a$ and $b$. The second largest eigenvalue $\lambda_2$ of the TM determines the largest correlation length of the system in the $\hat{x}$ direction.
}
  \label{fig:PepsMpsTM}
\end{figure}

\begin{figure}
  \begin{centering}
\includegraphics[width=8.0cm]{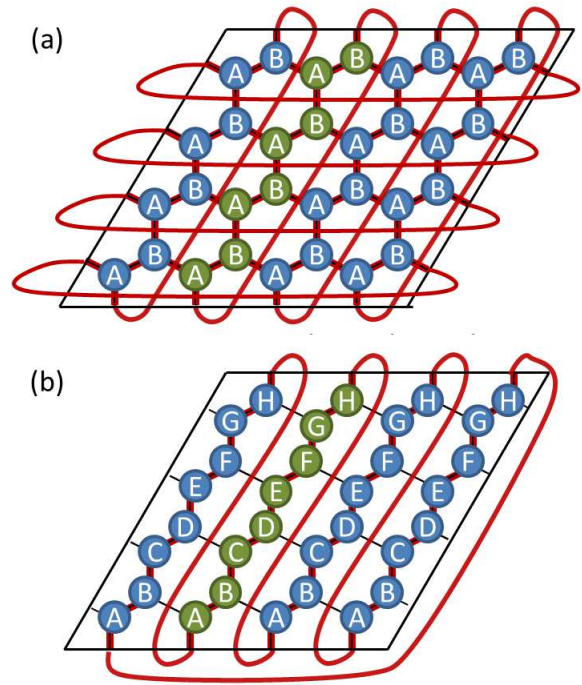}  \end{centering}
  \caption{%PepsMpsTor.eps
  (Color online) (a) PEPS representation for a state $\ket{\Psi^{\Tor}}$ on a finite torus of size $(L_x,L_y) = (4,4)$ obtained by reconnecting the PEPS tensors $A$ and $B$ that form a PEPS unit cell for a state $\ket{\Psi^{\Cyl}}$ on a cylinder of size $(L_x,L_y) = (\infty, 4)$. Notice that the resulting tensor network has the topology of a torus. (b) Similar construction for a MPS.
}
  \label{fig:PepsMpsTor}
\end{figure}

\subsection{Transfer matrix}

In the tensor network representation of a state of an infinite cylinder, correlations in the $\hat{x}$ direction can be seen to be mediated by powers of a transfer matrix (TM). This TM is expressed as a tensor network, see Fig. \ref{fig:PepsMpsTM}, made of reduced tensors. For instance, for a PEPS on a honeycomb lattice with tensors $A$ and $B$, the reduced tensors $a$ and $b$ are built from $A$ and $A^{*}$ (complex conjugate), and $B$ and $B^{*}$, respectively, according to
\begin{equation}\label{eq:a}
    (a)_{(\alpha \alpha')(\beta \beta')(\gamma \gamma')}\equiv \sum_{s} (A)_{\alpha \beta \gamma s}(A)^{*}_{\alpha' \beta' \gamma' s},
\end{equation}
where $(\alpha,\alpha')$ denotes a double index.

Let $\{\lambda_1, \lambda_2,\cdots \}$ denote the eigenvalues of the TM, ordered in deceasing absolute value, $|\lambda_1| \geq |\lambda_2| \geq \cdots$. In a normalized tensor network, the largest eigenvalue $\lambda_1$ of the TM is always one, $\lambda_1=1$, and the decay of correlators $c(x)$ in the $\hat{x}$ direction is governed by the rest of eigenvalues $\{\lambda_2, \lambda_3, \cdots\}$. In particular, the dominant correlations decay exponentially as
\begin{equation}\label{eq:xiTM}
    c(x) \approx (\lambda_2)^{x} = e^{-x/\xi_{\mbox{\tiny TN}}},
\end{equation}
where $\xi_{\mbox{\tiny TM}} \equiv -1/\log(\lambda_2)$ defines the TM correlation length. [Here we have assumed, for simplicity, that $|\lambda_2| > |\lambda_3|$; for $\lambda_2 = \lambda_3$ a similar expression holds].

When studying topological order with infinite cylinders of finite $L_y$, the TM correlation length $\xi_{\mbox{\tiny TM}}$ of the ground states offers a natural consistency check for the assumption that the value of $L_y$ is large enough to be representative of the thermodynamic limit. If the size $L_y$ of the cylinder is much larger than $\xi_{\mbox{\tiny TM}}$, then we indeed expect finite size effects to be very small, justifying a posteriori the use of a cylinder with that size $L_y$. If, on the contrary, $L_y$ is somewhat comparable to $\xi_{\mbox{\tiny TM}}$ (or $\xi_{\mbox{\tiny TM}}$ depends significantly on the size $L_y$ of the cylinder), it is likely that finite size effects are still dominant.

\subsection{From an infinite cylinder to a finite torus}

Given a tensor network representation of state $\ket{\Psi^{\Cyl}}$ on an infinite cylinder with finite size $L_y$ in the $\hat{y}$ direction, it is possible to obtain a tensor network representation of a state $\ket{\Psi^{\Tor}}$ on a torus with the same size $L_y$ and some finite size $L_x$ in the $\hat{x}$ direction by conveniently reconnecting into a torus the tensors in a proper subregion of the cylinder, as illustrated in Fig. \ref{fig:PepsMpsTor}. If in the cylinder we had  $\xi_{\mbox{\tiny TM}} \ll L_y$ for a ground state $\ket{\Psi^{\Cyl}_i}$ of $H$ with anyon flux $i$ in the $\hat{x}$ direction, then we expect that the resulting state $\ket{\Psi^{\Tor}_i}$ on a torus of size $(L_x,L_y)$ with $L_x \geq L_y$ be a good approximation to the ground state of $H$ on that torus with the same anyon flux $i$ in the $\hat{x}$ direction. The results of the present paper offer a clear confirmation of this expectation.
\begin{figure}
  \begin{centering}
\includegraphics[width=8.0cm]{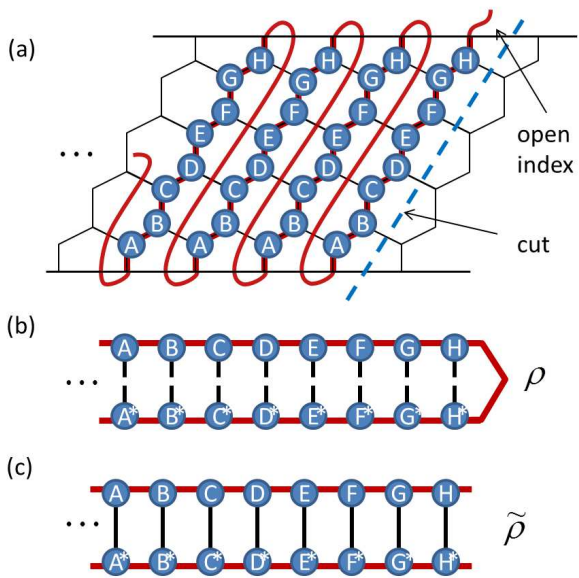}  \end{centering}
  \caption{%spectrumMPS.eps
 (Color online)
 (a)-(b) From an MPS for the state $\ket{\Psi^{\Cyl}}$ of an infinite cylinder, see Fig. \ref{fig:PepsMpsCyl}, a MPS for the reduced density matrix $\rho$ for half the infinite cylinder is obtain by tracing out the open index in this figure (assuming the proper normalization of the MPS tensors). (c) Alternatively, a density matrix $\tilde{\rho}$ with the same spectral properties as $\rho$ is obtained from this figure by tracing out all the physical indices (again, assuming the proper normalization of the MPS tensors) .
}
  \label{fig:spectrumMPS}
\end{figure}

\begin{figure}
  \begin{centering}
\includegraphics[width=8.0cm]{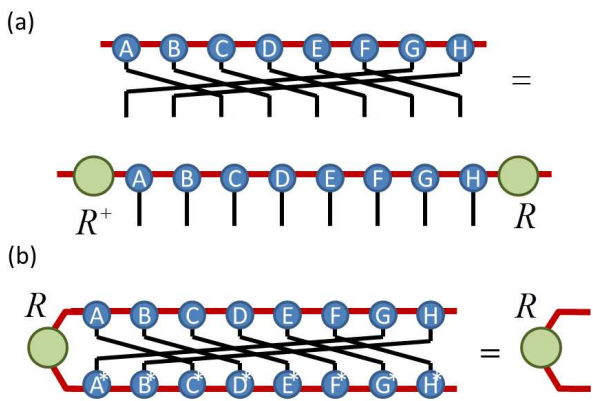}  \end{centering}
  \caption{%spectrumMPS2.eps
  (Color online)
  (a) A translation by one lattice unit cell (made of two sites) in the $\hat{y}$ direction is implemented by permuting the indices of the MPS unit cell (made of 8 tensors). If state $\ket{\Psi^{\Cyl}}$ is invariant under such a translation, then the MPS unit cell before and after the corresponding permutation of sites still describes the same state, and it is therfore related by a transformation $R$ that, under proper normalization of the MPS tensors, is unitary and applies a phase $e^{-i\frac{2\pi}{L_y}K}$ to each eigenvector of $\rho$, where $K=0,1,\cdots, L_y-1$ is the momentum of that eigenstate. (b) Matrix $R$ is obtained as the fixed point of a (mixed) transfer matrix.
}
  \label{fig:spectrumMPS2}
\end{figure}

\section{Entanglement spectrum of half an infinite cylinder from a matrix product state}

Given an infinite MPS for a state $\ket{\Psi^{\Cyl}}$ of an infinite cylinder, see Fig. \ref{fig:PepsMpsCyl}(b), we obtain a tensor network for the reduced density matrix $\rho$ for half of the cylinder by cutting the MPS into two halves, and then tracing over the open bond index, see Fig. \ref{fig:spectrumMPS}(a)-(b). [Here we assume that the open index is normalized as to contain the Schmidt coefficients of that cut]. Translations by one lattice unit cell (or two sites) in the $\hat{y}$ direction generate a finite Abelian group $\mathbb{Z}_{L_y}$ of transformations on the infinite cylinder. If $\ket{\Psi}$ is invariant under such translations, then so is $\rho$. Therefore each eigenvector of $\rho$ transforms according to some phase $e^{-i\frac{2\pi}{L_y}K}$, where $K=0,1,\cdots, L_y-1$ is the momentum of that eigenvector. Our goal is to obtain the spectrum $\{p_{\alpha}, K_{\alpha}\}$ of $\rho$.

If instead of tracing the open bond index, we trace out the infinitely many physical indices of half of the cylinder [now assuming that the open index does not contain the Schmidt coefficients of that cut], then we obtain a density matrix $\tilde{\rho}$, living on the vector space of the open bond index, with the same spectral properties as $\rho$, see Fig. \ref{fig:spectrumMPS}(c). We can then obtain the eigenvalues $\{p_{\alpha}\}$ of $\rho$ by diagonalizing $\tilde{\rho}$. In addition, we can also obtain the momentum $K_{\alpha}$ associated to each eigenvalue $p_{\alpha}$ by the construction in Fig. \ref{fig:spectrumMPS2}, where the matrix $R$ is a unitary representation of the Abelian group $\mathbb{Z}_{L_y}$ of translations in the $\hat{y}$ direction.

\section{Computation of quantum dimensions from Entanglement Entropy}

In this appendix we explain how to extract the quantum dimensions $\{d_i\}$ and total quantum dimension $D \equiv \sqrt{\sum_i (d_i)^2}$, from the ground states $\{\ket{\Psi_i^{\Cyl}}\}$ of a local Hamiltonian $H$ on an infinite cylinder of finite width $L_y$, where ground state $\ket{\Psi_i^{\Cyl}}$ has anyon charge $i$ inside. This approach is valid for any emergent anyon model. Recall that if anyon $i$ is Abelian, then $d_i=1$ (as is the case for the identity, $d_1=1$, which is present in any anyon model), whereas if anyon $i$ is non-Abelian, then $d_i >1$.

For a given ground state $\ket{\Psi_i^{\Cyl}}$, let $\rho_i$ denote the reduced density matrix for half of the infinite cylinder, and let $\{p_{i,\alpha}\}$ denote its eigenvalues. The TEE $\gamma_i$ for half of the infinite cylinder is then obtained from the scaling of the entanglement entropy $S(\rho_i) \equiv -\sum_\alpha p_{i,\alpha} \log(p_{i,\alpha})$ as a function of the width $L_y$ of the cylinder, which reads \cite{Kitaev06,Levin06}
\begin{equation}\label{eq:topoEntropy}
    S_i(L_y) = \alpha L - \gamma_i.
\end{equation}

Following Eq. 40 of Ref. \cite{Dong08} (see also Eq. 18 in Ref. \cite{Kitaev06}), the TEE of half a torus (divided appropriately through two cuts in the $\hat{y}$ direction) in the state $\ket{\Psi_i^{\Tor}}$ with anyon flux of type $i$ in the $\hat{x}$ direction is equal to $-2\log (d_i/D)$, where the factor $2$ comes from counting two uncorrelated boundaries. For half an infinite cylinder in state $\ket{\Psi_i^{\Cyl}}$ we then obtain half of this value, namely
\begin{equation}\label{eq:TEE}
    \gamma_i = -\log (d_i/D).
\end{equation}

We can use the numerical estimates of $\{\gamma_i\}$, and thus of $\{d_i/D\}$, for two purposes. First, it follows from the definition of $D\equiv \sqrt{\sum_i (d_i)^2}$ that $\sum_i (d_i/D)^2=1$. Therefore, if the sum $\sum_i (d_i/D)^2$ for our numerical estimates is close to one, we obtain a confirmation that the set of ground states $\ket{\Psi_i^{\Cyl}}$ is complete. If, on the contrary, this sum is significantly smaller than $1$, this may indicate that the set of ground states is not complete. Second, using that $d_1=1$ and $d_1\leq d_i$, we can use the smallest of the estimate $\{\gamma_i\}$ to obtain $d_1/D$, or $D$. Once we know the total quantum dimension $D$, we can extract each quantum dimension $d_i$ from $\gamma_i$.

Notice that $\alpha$ in Eq. \ref{eq:topoEntropy} is expected to be roughly the same for all ground states $\{\ket{\Psi_i^{\Cyl}}\}$. Indeed, $\alpha$ measures the local contributions to the entropy, and all the ground states are locally very similar, up to corrections that decay exponentially with $L_y$. Therefore, for a fixed value of $L_y$ (e.g. the largest affordable $L_y$ in a DMRG calculation, which is the one less affected by finite size effects) we can also compute $d_i/d_j$ by simply subtracting the entropies $S_i$ and $S_j$, since $S_i - S_j = -(\gamma_i-\gamma_j) = \log(d_i/d_j)$ . Then, using that the smallest quantum dimension is always $d_1=1$, we can compute the rest of quantum dimensions $d_i$ (and therefore also $D$). This alternative approach does not require an extrapolation in $L_y$. However, it does not allow us to check that the set of ground states $\{\ket{\Psi_i^{\Cyl}}\}$ is complete.

\section{Computation of the modular matrices $U$ and $S$ in a lattice with $\pi/3$ rotational symmetry}

 Let $\{\ket{\Psi_{i}^{\Tor}}\}$ be an orthogonal basis of (quasi-degenerate) ground states of some lattice Hamiltonian $H$ on a torus such that $H$ is invariant under a $\pi/3$ rotation $R_{\pi/3}$ (e.g. on a honeycomb, triangular or kagome lattice). In this appendix we show how to obtain the $U$ and $S$ modular matrices of the emergent anyon model from the matrix $V$ of scalar products between the basis $\{\ket{\Psi_i^{\Tor}}\}$ and the rotated basis $\{R_{\pi/3} \ket{\Psi_i^{\Tor}}\}$,
\begin{equation}\label{eq:V2}
    V_{ij} = \bra{\Psi_{i}^{\Tor}} R_{\pi/3} \ket{\Psi_j^{\Tor}}.
\end{equation}

Indeed, following Ref. \cite{Zhang12}, matrix $V$ corresponds to the product $D^{\dagger}US^{-1}D$, where $D$ is a diagonal matrix with phases $D_{jj} = e^{i\phi_j}$ (the $e^{i\phi_j}$ corresponds to an arbitrary phase in the definition of the ground state $\ket{\Psi_j^{\Tor}}$), $U$ is also diagonal with entries $U_{jj} = e^{-i\frac{2\pi}{24}c}\theta_j$ corresponding to the central charge $c$ of the anyon model and the twist $\theta_j$ of anyon type $j$, and $S$ has (in general complex) entries $S_{ij}$ with $S_{j1}=S_{1j} = d_j/D >0$. We therefore have
\begin{equation}\label{eq:V3}
    V_{ij} = e^{-i\frac{2\pi}{24}c} \theta_i (S_{ji})^{*}e^{-i(\phi_i-\phi_j)}.
\end{equation}
In particular, from
\begin{equation}\label{eq:V11}
    V_{11} = e^{-i\frac{2\pi}{24}c} \theta_1 (S_{11})^{*} = e^{-i\frac{2\pi}{24}c} S_{11},
\end{equation}
where we used that $\theta_1=1$ and $S_{11}>0$, we can read the central charge $c$ (modulus 24). Then, knowing $e^{-i\frac{2\pi}{24}c}$, from
\begin{eqnarray}\label{eq:V1j}
   \frac{V_{1j}}{e^{-i\frac{2\pi}{24}c}} = \theta_1(S_{j1})^{*} e^{-i(\phi_1-\phi_j)}
    =(S_{j1}) e^{-i(\phi_1-\phi_j)},
\end{eqnarray}
for $j>1$, where again we used that $\theta_1=1$ and $S_{j1}>0$, we can read the relative phases $e^{-i(\phi_1-\phi_j)}$, with which we can compute all $e^{-i(\phi_i-\phi_j)}$. Finally, from
\begin{equation}\label{eq:Vi1}
    \frac{V_{i1}}{e^{-i\frac{2\pi}{24}c} e^{-i(\phi_i-\phi_j)}} = \theta_i(S_{1i})^{*} = \theta_i S_{1i},
\end{equation}
for $i>1$, where we used that $S_{1i}>0$, we also learn the twist $\theta_i$. In this way, we have completely characterize the entries $U_{jj}$ of the modular transformation $U$. The entries $S_{ij}$ of the modular matrix $S$ are then obtained from
\begin{equation}\label{eq:Vij}
    \frac{V_{ij}}{ e^{-i\frac{2\pi}{24}c}\theta_i e^{-i(\phi_i-\phi_j)}} =(S_{ji})^{*},
\end{equation}
where we can use the fact that the exact $S$ matrix fulfils $S_{i1}=S_{1i}$ as a non-trivial check.

In the above derivation we assumed that we knew which state corresponds to identity charge $i=1$, and placed that state in the first position of rows and columns of various matrices. How can we know which ground state $\ket{\Psi_i^{\Cyl}}$ corresponds to the identity anyon charge? In a chiral theory, it might be possible to use the entanglement spectrum of the different ground states $\{\ket{\Psi_i^{\Cyl}}\}$ (or the doubled entanglement spectrum of $\ket{\Psi_i^{\Tor}}$) to identify the anyon charge $i=1$, for instance as the one containing the tower of the identity primary field of the corresponding CFT, as we did in the main text of this paper for the semion model. Alternatively, in some cases the identity charge may be the only anyon charge with quantum dimension $d_i=1$ (as computed e.g. through the entanglement entropy of $\ket{\Psi_i^{\Cyl}}$, see previous section, or directly from the entries of the $S$ matrix). If several states $\ket{\Psi_i^{\Tor}}$ have the same quantum dimension $d_i=1$ and nothing else allows to identify which one corresponds to the identity, then the above derivation can be applied repeatedly by assuming each time that a different charge with $d_i=1$ is the identity charge. From the resulting $S$ and $U$ matrices and the several non-trivial constraints they must fulfill, we may then be able to tell whether that assumption was correct, although we do not know if this will always succeed.

\begin{thebibliography}{99}

\bibitem{Wen90} %Topological Orders in Rigid States
X.-G. Wen, Int. J. Mod. Phys. B4, 239 (1990).

\bibitem{Laughlin83}
%Anomalous Quantum Hall Effect: An Incompressible Quantum Fluid with Fractionally Charged Excitations
R. B. Laughlin, Phys. Rev. Lett., 50, 1395 (1983).

\bibitem{Anderson73}
%Resonating valence bonds: A new kind of insulator?
P. W. Anderson, Mater. Res. Bull. 8 (2) (1973).

\bibitem{Balents10}
%Spin liquids in frustrated magnets
L. Balents, Nature 464 (7286): 199"1¤78 (2010).

\bibitem{Tsui82}
%Two-Dimensional Magnetotransport in the Extreme Quantum Limit
%D. C. Tsui, H. L. Stormer, A. C. Gossard,
D. C. Tsui \textit{et al}.,
Phys. Rev. Lett., 48, 1559 (1982).

\bibitem{Bednorz86}
%Possible high TC superconductivity in the Ba-La-Cu-O system
%J. G. Bednorz, K. A. Mueller,
J. G. Bednorz, \textit{et al}.,
%Zeitschrift fuer Physik
Z. Phys. B 64 (2): 189"1¤73 (1986).

\bibitem{Anderson87}
%The resonating valence bond state in La2CuO4 and superconductivity
P. W. Anderson, Science 235 (4793): 1196-1198 (1987).

\bibitem{Wen89}
%Chiral Spin States and Superconductivity
%X.-G. Wen, F. Wilczek, A. Zee,
X.-G. Wen \textit{et al}.,
Phys. Rev., B39, 11413 (1989).

\bibitem{Kitaev03}
% Fault-tolerant quantum computation by anyons
A. Y. Kitaev, Annals Phys. 303, 2-30(2003).

\bibitem{Nayak07}
%Non-Abelian Anyons and Topological Quantum Computation
%C. Nayak, S. H. Simon, A. Stern, M. Freedman, S. Das Sarma,
C. Nayak \textit{et al}.,
Rev. Mod. Phys. 80, 1083 (2008).

\bibitem{Kitaev06}
%Topological Entanglement Entropy
A. Kitaev, J. Preskill, Phys. Rev. Lett. 96, 110404 (2006).

\bibitem{Levin06}
%Detecting Topological Order in a Ground State Wave Function
M. Levin, X.-G. Wen, Phys. Rev. Lett. 96, 110405 (2006).

\bibitem{Li08}
%Entanglement Spectrum as a Generalization of Entanglement Entropy: Identification of Topological Order in Non-Abelian Fractional Quantum Hall Effect States
%H. Li, F. D. M. Haldane,
H. Li \textit{et al}.,
 Phys. Rev. Lett. 101, 010504 (2008).

\bibitem{Zhang12}
% Quasi-particle Statistics and Braiding from Ground State Entanglement
%Y. Zhang, T. Grover, A. Turner, M. Oshikawa, A. Vishwanath,
Y. Zhang \textit{et al}.,
Phys. Rev. B 85, 235151 (2012).

\bibitem{Grover11}
% Quantum Entanglement and Detection of Topological Order in Numerics
T. Grover, arXiv:1112.2215v1

\bibitem{White92}
% DMRG
S. R. White, Phys. Rev. Lett. 69, 2863 (1992).
%S. R. White, Phys. Rev. B. 48, 10345 (1993).

\bibitem{Verstraete04}
%Renormalization algorithms for Quantum-Many Body Systems in two and higher dimensions
F. Verstraete, J. I. Cirac, arXiv:cond-mat/0407066v1 %[cond-mat.str-el]

\bibitem{Vidal07}
% Entanglement Renormalization
G. Vidal, Phys. Rev. Lett. 99, 220405 (2007).

\bibitem{Jordan08}
%Classical simulation of infinite-size quantum lattice systems in two spatial dimensions
%J. Jordan, R. Orus, G. Vidal, F. Verstraete, J. I. Cirac,
J. Jordan, \textit{et al}.,
Phys. Rev. Let. 101, 250602 (2008).

\bibitem{Aguado08}
% Entanglement Renormalization and Topological order
%M. Aguado, G. Vidal,
M. Aguado \textit{et al}., Phys. Rev. Lett. 100, 070404 (2008).

\bibitem{Yan11}
%Spin Liquid Ground State of the $S=1/2$ Kagome Heisenberg Model
%S. Yan, D. A. Huse, S. R. White,
S. Yan \textit{et al}.,
Science, Vol. 332 (6034) 1173-1176 (2011).

\bibitem{Jiang12}
%Identifying Topological Order by Entanglement Entropy
%H.-C. Jiang, Z. Wang, L. Balents,
H.-C. Jiang, \textit{et al}. arXiv:1205.4289v1 %[cond-mat.str-el]

\bibitem{Depenbrock12}
%Nature of the Spin Liquid Ground State of the S=1/2 Kagome Heisenberg Model
%S. Depenbrock, I. P. McCulloch, U. Schollwoeck,
S. Depenbrock \textit{et al}.,
arXiv:1205.4858v1 %[cond-mat.str-el]

\bibitem{Haldane88}
%Model for a quantum hall effect without landau levels: Condensed-matter realization of the "parity anomaly"1¤7
F. D. M. Haldane, Phys. Rev. Lett. 61, 2015 (1988).

\bibitem{Wang11}
%Fractional quantum hall effect of hard-core bosons in topological flat bands
%Y.-F. Wang, Z.-C. Gu, C.-D. Gong, D. N. Sheng, Y.-F.
Wang \textit{et al}.,
Phys. Rev. Lett. 107, 146803 (2011).

\bibitem{Bonderson07} P. Bonderson, \textit{Non-Abelian anyons and interferometry}, PhD Thesis Caltech (2007).

\bibitem{DiFrancesco}
%P. Di Francesco, P. Mathieu, and D. Senechal,
P. Di Francesco \textit{et al}.,
\textit{Conformal Field Theory} (Springer, 1997).

\bibitem{Lukasz12}
L. Cincio \textit{et al}., in preparation.

\bibitem{McCulloch08}
% Infinite size density matrix renormalization group, revisited
I. P. McCulloch, arXiv:0804.2509v1 %[cond-mat.str-e1]

\bibitem{Crosswhite08}
% Applying matrix product operators to model systems with long-range interactions
%G. M. Crosswhite, A.C. Doherty and G. Vidal,
G. M. Crosswhite \textit{et al}.,
Phys. Rev. B 78, 035116 (2008).

\bibitem{Dong08}
%Topological entanglement entropy in Chern-Simons theories and quantum Hall fluids
%S. Dong, E. Fradkin, R. G. Leigh, S. Nowling,
S. Dong \textit{et al}.,
JHEP 0805, 016 (2008).

\bibitem{Hastings05}
%Quasi-adiabatic Continuation of Quantum States: The Stability of Topological Ground State Degeneracy and Emergent Gauge Invariance
%M. B. Hastings, X.-G, Wen,
M. B. Hastings \textit{et al}.,
Phys. Rev. B72 (2005) 045141.

\bibitem{Note} The states $\{\ket{\Psi_i^{\Tor}}\}$ correspond to the \textit{minimum entropy states}
MESs of Ref. \cite{Zhang12}. There, a complete basis of the ground subspace and a minimization over entanglement entropy is required in order to obtain the MESs. Instead, our approach only requires a lattice Hamiltonian $H$ realizing the topologically ordered phase.

\bibitem{Verlinde}
% Fusion rules and modular transformations in 2D conformal field theory
E. Verlinde, Nucl. Phys. B 300 (3): 360"1¤76 (1988).

\bibitem{Sandvik07}
% Variational Quantum Monte Carlo Simulations with Tensor-Network States
%A. W. Sandvik, G. Vidal,
A. W. Sandvik \textit{et al}.,
Phys. Rev. Lett. 99, 220602 (2007).


\end{thebibliography}
\end{document}